# Effect of Zn doping on the Magneto-Caloric effect and Critical Constants of Mott Insulator $MnV_2O_4$

Prashant Shahi[1], Harishchandra Singh[2], A. Kumar[1], K. K. Shukla[1], A. K. Ghosh[3], A.K. Yadav[4], A. K. Nigam[5], and Sandip Chatterjee[1,*]

*[1)] Department of Applied Physics, Indian Institute of Technology (Banaras Hindu University), Varanasi-221005, India*

*[2)] Indus Synchrotron Utilization Division, Raja Ramanna Centre for Advanced Technology, Indore - 452013, India*

*[3)] Department of Physics, Banaras Hindu University, Varanasi-221005, India*

*[4)] Atomic & Molecular Physics Division, Bhabha Atomic Research Centre, Mumbai – 400 094, India*

*[5)] Department of CMP & MS, Tata Institute of Fundamental Research, Mumbai-400005, India*

**Abstract**

X-ray absorption near edge spectra (XANES) and magnetization of Zn doped $MnV_2O_4$ have been measured and from the magnetic measurement the critical exponents and magnetocaloric effect have been estimated. The XANES study indicates that Zn doping does not change the valence states in Mn and V. It has been shown that the obtained values of critical exponents $\beta$, $\gamma$ and $\delta$ do not belong to universal class and the values are in between the 3D Heisenberg model and the mean field interaction model. The magnetization data follow the scaling equation and collapse into two branches indicating that the calculated critical exponents and critical temperature are unambiguous and intrinsic to the system. All the samples show large magneto-caloric effect. The second peak in magneto-caloric curve of $Mn_{0.95}Zn_{0.05}V_2O_4$ is due to the strong coupling between orbital and spin degrees of freedom. But 10% Zn doping reduces the residual spins on the V-V pairs resulting the decrease of coupling between orbital and spin degrees of freedom.

## Introduction

The magnetocaloric effect (MCE) has drawn great attention as it provides a technique for studying magnetic phase transition and also for commercial applications, since it has a potential application in magnetic refrigeration.[1–3] One of the important factors for using magnetic refrigeration is to find a suitable material within the working temperature range of refrigeration. This accounts for the recent increasing interest in the study of various physical properties characterizing magnetic materials that show large entropy changes when subjected to the magnetic intensity variation. Generally, the magnetic transitions are of second order type for the materials exhibit conventional MCE, and the changes to the magnetic contribution to entropy is only of magnetic origin.[4] Moreover, MCE involving a first order magnetic phase transition has also attracted much interest.[5] When a first order magnetic transition occurs, MCE is a giant one and is enriched by considerable contribution from the lattice through latent heat.[6-9]

Recently, in spinel ferromagnetic $MnV_2O_4$ large magneto-caloric effect is reported around $T_c$ (=57K).[10] $MnV_2O_4$ is the member of the Vanadium oxide spinels family which shows the orbital degeneracy, and the interplay of spin, orbital and lattice degrees of freedom. Other than the conventional spin-orbit coupling this interplay also arises due to the preference of the occupation of the specific orbitals with geometrical anisotropy with specific types of magnetic interaction in specific direction (the so-called Kugel-Khomskii type coupling).[11] Moreover, in these materials a sequence of two phase transitions are obserevd.[12-14] Recent attention has turned to $MnV_2O_4$.[15-18] This $MnV_2O_4$ lies in the Mott insulator regime. Also, in this case $Mn^{2+}$ is in the $3d^5$ high-spin (with S=5/2) configuration without any orbital degrees of freedom. Furthermore, the B-site is occupied by the $V^{3+}$ ion, which takes the $3d^2$ high-spin configuration in the triply degenerate $t_{2g}$ orbital, and has orbital degrees of freedom. $MnV_2O_4$ exhibits a collinear ferrimagnetic ordering at $T_c$=57K (second order transition), where the magnetic moments of the Mn and the V sites align to the opposite direction, and then structural phase transition from a cubic to a tetragonal phase at $T_s$=53K (first order transition), with the spin structure becoming non-collinear.[15] It was also found that the cubic to tetragonal transition could be induced by a magnetic field of few tesla.[16.17] Moreover, the orbital state of $MnV_2O_4$ cannot be explained simply by anti-ferro orbital model.[19] The large magneto-caloric effect in this compound is suggested to be related to the orbital entropy change due to the change of the orbital state of $V^{3+}$ ions with an applied field around $T_c$ (=57K).[10] Furthermore, when Zn is doped on the Mn site the value of magneto-caloric effect increases and the maximum is observed at the lower transition temperature ($T_S$) which has been attributed to orbital ordering.[20] Zn doping also revealed that the structural phase transition from a cubic to a tetragonal phase is a co-operative phenomena dominated by orbital degrees of freedom.[16] In this perspective the detail study of Zn doping in $MnV_2O_4$ will be highly interesting to throw light on the mechanism of $MnV_2O_4$. Moreover, the critical behaviour, which has direct correlation with the MCE, of $MnV_2O_4$ might also be distinctive to provide interesting information about the magnetic spin ordering in this spinel Vanadate. Baek et al.[18] have reported the critical exponents of this system. From the ac-susceptibility measurement they obtained very unusal values of the critical constants ($\beta$ =0.36 and $\gamma$=0.59).[18] In a recent paper Zhang et al. reported the critical constants $\beta$ ~0.349 (close to 3D Heisenberg model) and $\gamma$~0.909 (close to mean field model).[21] Moreover, Garlea et al. determined the $\beta$ value from the integrated intensity I(T) fitting.[22] The I(T) can be described near the two phase transitions as $I(T) \propto (T_{C,S}-T)^{2\beta}$. The obtained value by them was close to 3D Heisenberg and 3D Ising models. To resolve the issue we have also studied the variation of critical constants with Zn doping using modified Arrot plots.

**Experimental**

The polycrystalline $Mn_{1-x}Zn_xV_2O_4$ samples used in this study were prepared by solid state reaction method. Appropriate ratio of MnO, ZnO and $V_2O_3$ were ground thoroughly and pressed into pellets. The pellets were sealed in evacuated quartz tube and heated at 950$^o$C for 40 hours. The X-ray powder diffraction (XRD) has been taken from Rigaku MiniFlex II DEXTOP X-ray Diffractometer with Cu-kα radiation. The XANES measurements have been carried out at the Energy-Scanning EXAFS beamline (BL-9) in transmission mode at the INDUS-2 Synchrotron Source (2.5 GeV, 100 mA) at Raja Ramanna Centre for Advanced Technology (RRCAT), Indore, India. This beamline operates in energy range of 4 KeV to 25 KeV. The beamline optics consist of a Rh/Pt coated collimating meridional cylindrical mirror and the collimated beam reflected by the mirror is monochromatized by a Si(111) (2d=6.2709) based double crystal monochromator. The second crystal of DCM is a sagittal cylinder used for horizontal focusing. Three ionization chambers (300 mm length each) have been used for data collection in transmission mode, one ionization chamber for measuring incident flux ($I_0$), second one for measuring transmitted flux ($I_t$) and the third ionization chamber for measuring EXAFS spectrum of a reference metal foil for energy calibration. Magnetic measurement was done using Magnetic Properties Measurement System (MPMS) SQUID (Quantum Design) magnetometer with the bulk samples. Data were collected upon warming up the sample.

**Results and Discussion**

Fig. 1 shows the XRD patterns of $Mn_{1-x}Zn_xV_2O_4$ indicating the single phase of all the samples. We have refined the diffraction data with the Reitveld refinement program (the fitted curve has also been shown in Fig.1) the fitted parameters are shown in Table-I. The lattice parameter (a) measured from the XRD data (using Reitveld refinement) is plotted in the inset of Figure 1. From the figure, it is clear that the increase in Zn doping has decreased the size of lattice parameters, which is attributed to the ionic size mismatch between $Mn^{2+}$(0.80 Å) and $Zn^{2+}$(0.74 Å). The linear dependence of the lattice constant on Zn concentration indicates that the doping of Zn ion does not change the structure of $MnV_2O_4$. The samples are homogeneous solid solutions and Zn is doped into the crystal lattice following the Vegard's law,[23] indicating that the $Zn^{2+}$ replaces well the $Mn^{2+}$. Furthermore, because no diffraction peaks corresponding to Zn or Zn-related impurity phases are observed, the formation of $Mn_{1-x}Zn_xV_2O_4$ solid solution is confirmed instead of any Zn precipitation or second phase. In order to probe the oxidation state of all the transition metals i.e. Mn, Zn and V, we have performed XANES measurements at their corresponding K edges. Figure 2 shows the edge step normalized XANES spectra for Mn K-edge for the $Mn_{1-x}Zn_xV_2O_4$ ($x$ = 0, 0.05 & 0.1) samples. The XANES spectra of all the samples are plotted with three standards Mn metal foil, MnO and $Mn_2O_3$ with +0, +2 and +3 oxidation state respectively. A typical K edge XANES spectra exhibits structured pre-edge region and the dominant peak called white line peak followed by main rising edge.[24] Here, we emphasize only the main edge part of the XANES spectra for our purpose. The pre-edge features below the main edge are due to Mn 1*s* transition into unoccupied O 2*p*-Mn 3*d* (or Mn 3*d*/4*p*) hybridized states, which have *p* components projected at the Mn site as observed in many transition-metal oxides. The main edge feature at the Mn *K*-edge corresponds to the high-energy Mn 4*p* states. Energy position of the main edge in XANES spectra of a sample may be determined either as the energy corresponding to ~0.5 absorption or maximum energy value of a first order differentiated spectrum.[24] The main edge position of $Mn_{1-x}Zn_xV_2O_4$ for $x$ = 0, 0.05 and 0.1 samples coincides with MnO, for which Mn ion has been assigned to have a valence or effective charge of +2.

Figure 3 shows the edge step normalized XANES spectra for Zn K-edge for the $Mn_{1-x}Zn_xV_2O_4$ for $x$ = 0.05 and 0.1 along with two standards namely Zn metal and ZnO with +0 and +2 oxidation state respectively. This edge position in $Mn_{1-x}Zn_xV_2O_4$ for $x$ = 0.05 & 0.1 clearly shows that Zn is present in +2 oxidation state as its edge coincide with ZnO edge.

Figure 4 shows the same edge step normalized XANES spectra for V K-edge for the $Mn_{1-x}Zn_xV_2O_4$ ($x$ = 0, 0.05 and 0.1) along with standards V metal and $VOSO_4$ with +0 and +4 oxidation state respectively. A visual inspection of Figure 3, the main edge energy corresponding to V K edge is close to +4 but very far from +0. Energy positions of these samples, discussed above, well matches with the earlier reports for $V^{3+}$ [25]. A table for their energy positions with the corresponding references is presented in Table 2, indicating +3 oxidation state of Vanadium in our studied samples. From the above discussion it is clear that no change in the valence states of Mn and V is occurring due to doping of Zn.

Fig. 5 shows the temperature dependence of magnetization of $Mn_{1-x}Zn_xV_2O_4$ under zero field cooled (ZFC) condition at 100 Oe. The M-T curve of $MnV_2O_4$ exhibits a sharp paramagnetic-ferrimagnetic (PM-FM) phase transition. It is observed that the magnetization drops sharply for all the Zn doped $MnV_2O_4$ samples on cooling. This is due to the spin-pairing of V-V bonds.[26] Inset of Fig. 5 shows the evolution of $T_C$ with the V-V separation ($R_{V-V}$), determined from the Reitvelt analysis, for the $Mn_{1-x}Zn_xV_2O_4$ spinels. With decrease of the $R_{V-V}$ the overlap integral in t of $J \propto t^2/U$ [J is the super exchange spin-spin interaction] increases, the energy U must decrease or remain constant. As a matter of fact, within the localized-electron limit, $T_C$ should increase with decreasing V-V separation. In the present investigation it has been observed when Zn is doped V-V separation decreases, but also $T_c$ decreases. This suggests that the energy U in the localized electron superexchange theory is not constant when Zn is doped.

The M(H) curves at different temperatures (around the $T_c$) have been shown in Fig.6. According to the Scaling hypothesis,[27] a second order magnetic phase transition near the Curie point is characterized by a set of critical exponents of $\beta$, $\gamma$ and $\delta$ and the magnetic ordering can be studied

$$(H/M)^{1/\gamma} = C_1(T-T_c) + C_2M^{1/\beta} \quad (1)$$

which combines the relations for the spontaneous magnetization below $T_c$

$$M\sim (T_c-T),$$

and the inverse magnetic susceptibility above $T_c$

$$\chi^{-1} \sim (T-T_c)$$

To find the correct values of β and γ, linear fits to the isotherms are made to get the intercepts giving M(T) and χ(T). These new values of β and γ are then used to make a new modified Arrott plot. New values of critical exponents thus obtained are re-introduced in the scaling of the modified Arrott plot. The process is repeated until the iteration converges, leading to an optimum fitting value.

Fig. 7 shows the final result for the $Mn_{1-x}Zn_xV_2O_4$ samples. We have taken M(H) isotherms from 50K to 70K in every 2K interval. The calculated values of β and γ are, respectively, 0.393 and 1.01 for x=0 sample. For x=0.05 the values are respectively, 0.40 and 1.02 whereas for x=0.1 the values of β and γ are respectively, 0.42 and 1.07. Below 54 K it deviates from linearity due to the first order transition at ~52K.

The critical values we obtained are not from the universality class. According to the scaling theory the magnetization equation can be written as $M(H,\varepsilon)\varepsilon^{-\beta} = f_\pm(H/\varepsilon^{\beta+\gamma})$, where ε is the reduced temperature, $f_+$ for $T > T_c$ and $f_-$ for $T< T_c$ are regular functions.[28] The equation states that the plot between $M|\varepsilon|^{-\beta}$ vs $H|\varepsilon|^{-(\beta+\gamma)}$ gives two universal curves: one for $T > T_c$ and other for $T < T_c$. As shown in Fig. 8, the curves are divided in two parts one above $T_c$ and one below $T_c$. The inset of Fig. 8 also shows log-log plot and this also falls into two classes one above $T_c$ and one below $T_c$, in agreement with the scaling theory. Therefore the FM behaviour around Curie temperature get renormalized following the scaling equation of state indicating that the calculated critical exponents are reliable. Moreover, exponents often show various systematic trends or crossover phenomenon's one approaches $T_c$.[29,30] This occurs due to the presence of various competing couplings and/or disorder. For this reason, it is useful to introduce temperature-dependent effective exponents for ε≠0. It can be mentioned that effective exponents are non universal properties, and they are defined as:

$$\beta^{eff}(\varepsilon) = \frac{d[\ln Ms(\varepsilon)]}{d[ln\varepsilon]} \,. \quad \gamma^{eff}(\varepsilon) = \frac{d[ln\chi_0^{-1}(\varepsilon)]}{d[ln\varepsilon]} \quad (2)$$

We have calculated the $\beta^{eff}$ and $\gamma^{eff}$ by using the Eq. 2 (not shown), which do not match with universality class.

It is observed from the above discussion that the critical exponents of the present investigated sample are not consistent with the universality class. The similar behaviour has been observed in perovskite $Pr_{0.5}Sr_{0.5}MnO_3$ ($\beta$ =0.397and $\gamma$=1.331)[31] and in $La_{0.7}Sr_{0.3}MnO_3$ ($\beta$ =0.45 and $\gamma$=1.2)[32] due to phase separation. Other than perovskite, $Gd_{80}Au_{20}$ also shows unusual critical exponents ($\beta$ =0.44(2) and $\gamma$=1.29(5)) arise due to the dilution of global spin with the substitution of non-magnetic ions.[33] In the present case very close to the second order PM-FM transition (at~58K) there exists a first order structural transition (~52K) which is associated with the collinear to non-collinear spin transition. This may cause a large spin fluctuation which may be responsible for the unusual critical exponents between the actual material and the theoretical model.

In order to further investigate we have also estimated the magneto caloric effect of all the samples. The magnetic entropy change is given by

$$|\Delta Sm| = \sum_i \frac{Mi-Mi+1}{Ti+1-Ti} \Delta Hi \qquad (3)$$

Where $M_i$ is the Magnetization at Temperature $T_i$.[34] The obtained $|\Delta S_m|$ has been plotted as a function of temperature in Fig 9. In the inset of Fig. 6 the magnetic field variation of $|\Delta S_m|$ shows the $H^{2/3}$ dependency (the value of the exponent for x=0, 0.05 and 0.1 are respectively, 0.633, 0.67 and 0.666). This is consistent with the relation between magnetic entropy and the magnetic field near the magnetic phase transition which is given by[35]

$$|\Delta S_m| = -1.07qR(g\mu_B JH/kT_c)^{2/3} \qquad (4)$$

where q is the number of magnetic ions, R is the gas constant, and g is the Landau factor. In Fig. 9 it is observed that as Zn content increases the magnetic entropy value decreases, which is also consistent with the Eq. 4. As with increase of Zn content the magnetic Mn ions decrease. Moreover, in $Mn_{1-x}Zn_xV_2O_4$ for x=0 and 0.1 only one peak is found and that is at $T_c$ but for x=0.05 two peaks are observed (one is at $T_c$ and another is at $T_{oo}$). The observed behaviour for x=0 and x=0.05 is consistent with those reported.[10,20] It has been explained that the large magneto caloric effect in $MnV_2O_4$ is due to the change of the orbital state of $V^{3+}$ ions with applied field around $T_c$ which leads to the change in orbital entropy.[10] The observed second peak in x=0.05 is suggested to be due to the strong coupling between orbital and spin degrees of freedom.[20] But in that case in $MnV_2O_4$ sample also we should get two peaks. Moreover, in the present investigation x=0.1 sample does not show second peak. It might be the fact that in $MnV_2O_4$ the two transitions are very close to each other and because of that two peaks overlap into a single peak. Moreover for 10% Zn doping the chemical pressure increases which reduces the residual spins on the V-V pairs which leads to the decrease of long range magnetic ordering. As a matter of fact the coupling between orbital and spin degrees of freedom decreases. This might be the reason of diminishing the peak at low temperature when 10% Zn is doped

## Conclusion

The XANES study indicates that no change in the valence states of Mn and V is occurring due to doping of Zn and V remains in 3+ state. The giant magneto-caloric effect value is observed in these spinel vanadates and the entropy change (MCE value) decreases with increase of Zn content  It has been shown that the obtained values of the critical exponents $\beta$, $\gamma$ and $\delta$ do not belong to universal class and the values are in between the 3D Heisenberg model and mean field interaction model. The magnetization data follow the scaling equation and collapse into two branches indicating that the calculated critical exponents and critical temperatures are unambiguous and intrinsic to the system. The observed double peaks in MCE of $Mn_{0.95}Zn_{0.05}V_2O_4$ are due to the strong coupling between orbital and spin degrees of freedom. In this composition (x=0.05 ) the orbital ordering becomes maximum which in effect increase the coupling between orbital and spin degrees of freedom at $T_{oo}$ leading the second peak in magneto-caloric behavior in x=0.05 sample . When Zn content increases (*viz*. x=0.1) the chemical pressure increases which reduces the residual spins on the V-V pairs which leads to the decrease of long range magnetic ordering. As a consequence coupling between orbital and spin degrees of freedom decreases.

## Acknowledgements

SC is grateful to the funding agencies DST (Grant No.: SR/S2/CMP-26/2008) and CSIR (Grant No.: 03(1142)/09/EMR-II) and BRNS, DAE (Grant No.: 2013/37P/43/BRNS) for financial support. PS is grateful to CSIR, India for providing Research Fellowship. Authors are also grateful to D. Budhikot for his help in magnetization measurement.

**References:**

*Corresponding author e-mail id: schatterji.app@iitbhu.ac.in

**Table 1** Structural parameters (lattice parameters, V-V bond lengths) of $Mn_{1-x}Zn_xV_2O_4$ (with x=0, 0.05, 0.1) samples obtained from Reitveld refinement of X-Ray Diffraction data. The structural data have been refined with Space group Fd-3m at 300K.

| | X=0.0 | X=0.05 | X=0.1 |
|---|---|---|---|
| a(Å) | 8.5220(7) | 8.5150(3) | 8.5092(9) |
| V- V (Å) | 3.0130 (1) | 3.0105 (1) | 3.0105 (1) |

**Table-2**

| **Vanadium 1st derivative spectra peak energies** | | | | | | |
|---|---|---|---|---|---|---|
| V metal | V metal[3] | $V^{3+}{}_2O_3$[3] | *x* = 0 | *x* = 0.05 | *x* = 0.10 | $V^{4+}OSO_4$ |
| 5464.60 | 5465 | 5475.7 | 5475.37 | 5475.32 | 5475.34 | 5478.13 |

**Figure Captions:**

1. The X-ray diffraction pattern with the Reitveld refinement for the $Mn_{1-x}Zn_xV_2O_4$ (with x = 0.0, 0.05.0.1). The inset shows the variation of lattice parameters with Zn concentration. (Colour online)
2. Normalised XANES spectra of $Mn_{1-x}Zn_xV_2O_4$ for $x$ = 0, 0.05 & 0.1 at Mn K-edge with along with reference Mn metal, MnO and $Mn_2O_3$ sample. (Colour online)
3. Normalised XANES spectra of $Mn_{1-x}Zn_xV_2O_4$ for $x$ = 0.05 & 0.1 at Zn K-edge with along with Zn metal and ZnO standard sample. (Colour online)
4. Normalised XANES spectra of $Mn_{1-x}Zn_xV_2O_4$ for $x$ = 0, 0.05 & 0.1 at V K-edge with along with V metal and VOSO4 standard sample. (Colour online)
5. The temperature variation of magnetization of $Mn_{1-x}Zn_xV_2O_4$ at 100 Oe magnetic field. Inset showing the $T_N$ (up arrow) and $T_S$ (down arrow) of all the samples.
6. Magnetization as a function of applied magnetic field for the $Mn_{1-x}Zn_xV_2O_4$ (with x = 0.0, 0.05.0.1) at different temperatures (Isotherms have been measured every 2 K interval around Curie temperature). (Colour online)
7. Final results for critical constants of $Mn_{1-x}Zn_xV_2O_4$ (with x=0, 0.05, 0.1). (colour online)
8. Universal curves and inset shows the log–log plot of universal curves of $Mn_{1-x}Zn_xV_2O_4$ (with x=0, 0.05, 0.1). (Colour online)
9. Magneto caloric effect of $Mn_{1-x}Zn_xV_2O_4$ at 2T and 4T magnetic fields. Inset shows the fitting of $\Delta S$ (entropy change) vs magnetic field (H) curve of $Mn_{1-x}Zn_xV_2O_4$ (with x=0, 0.05, 0.1). (Colour online)

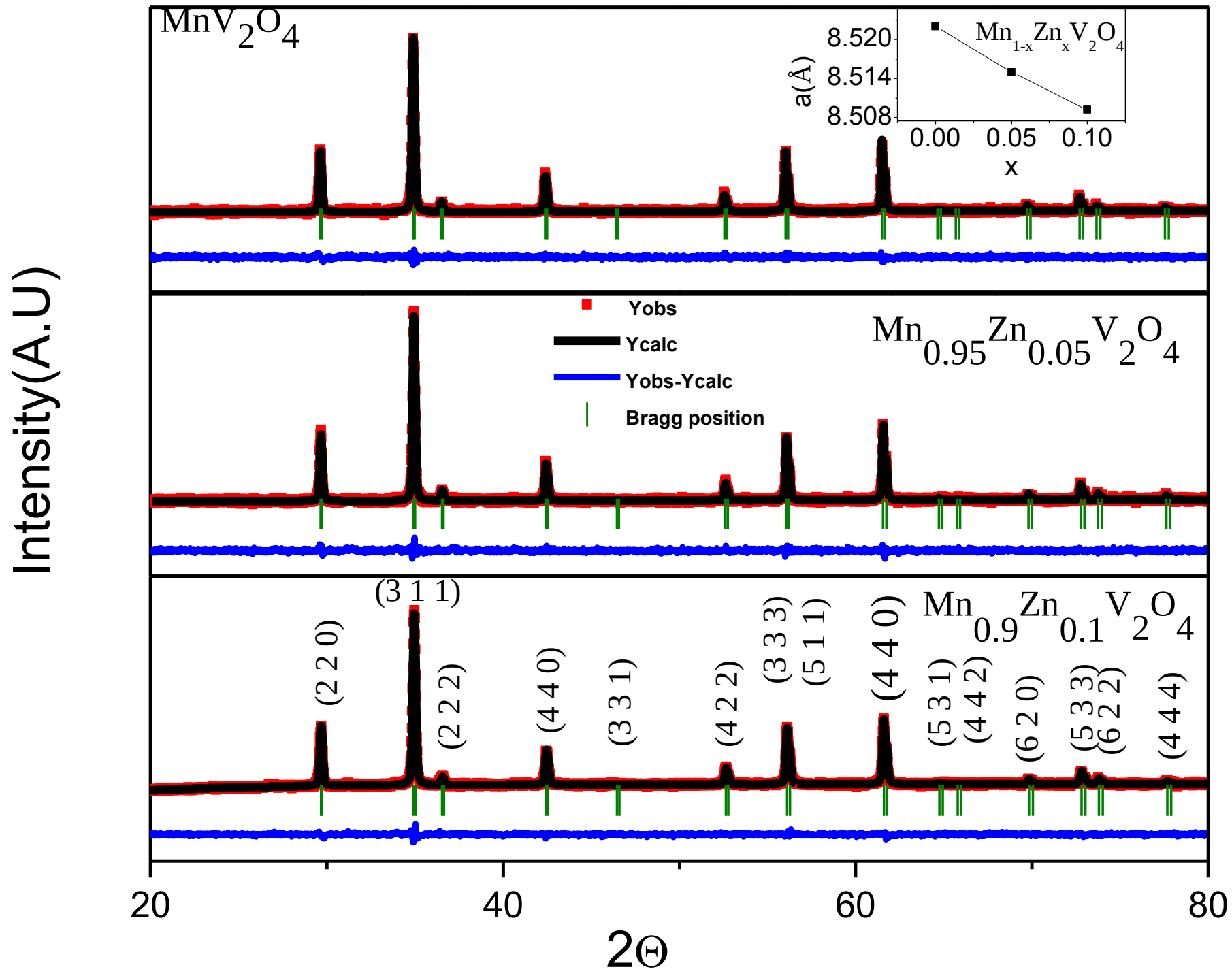

FIG.1

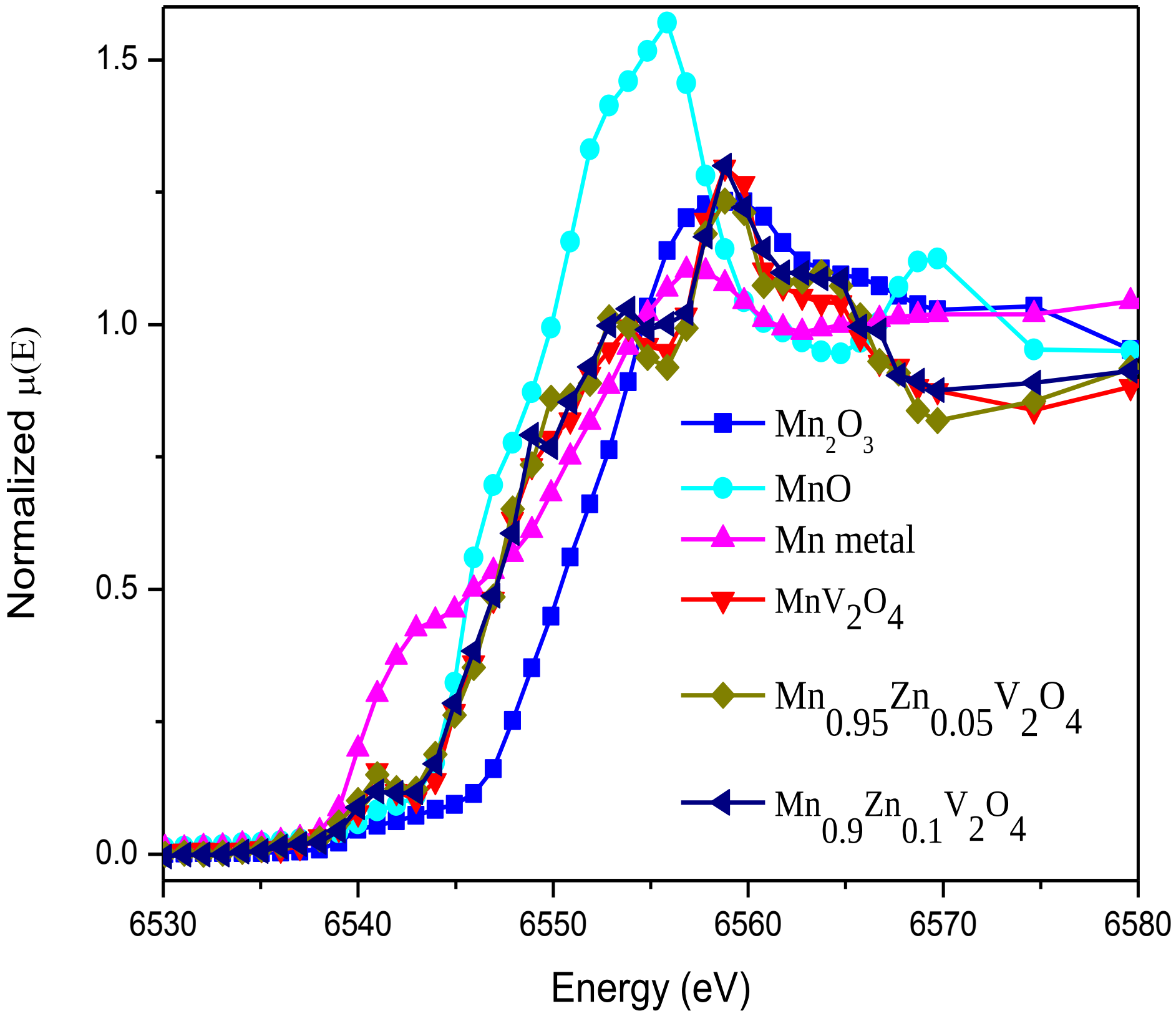

FIG.2

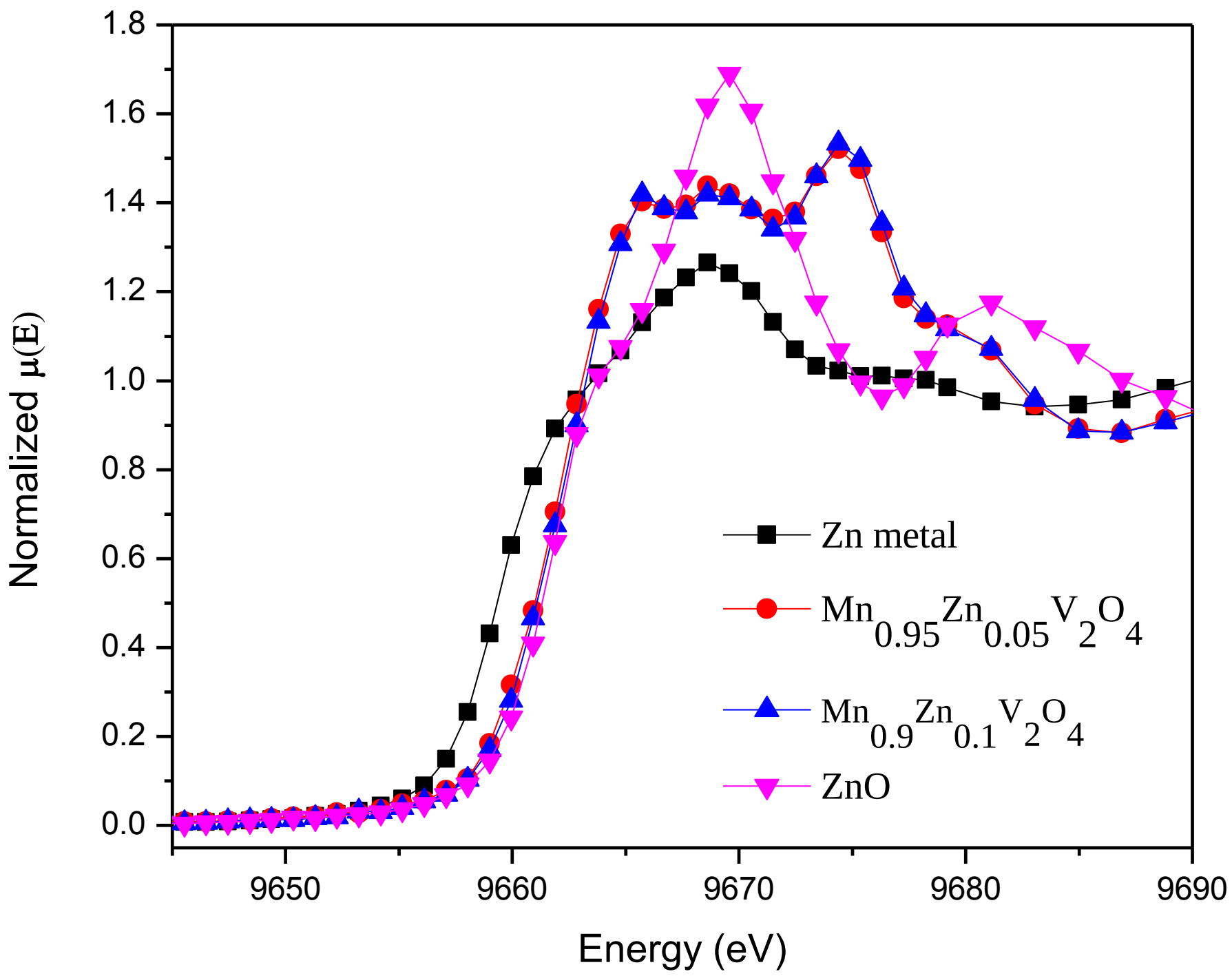

FIG.3

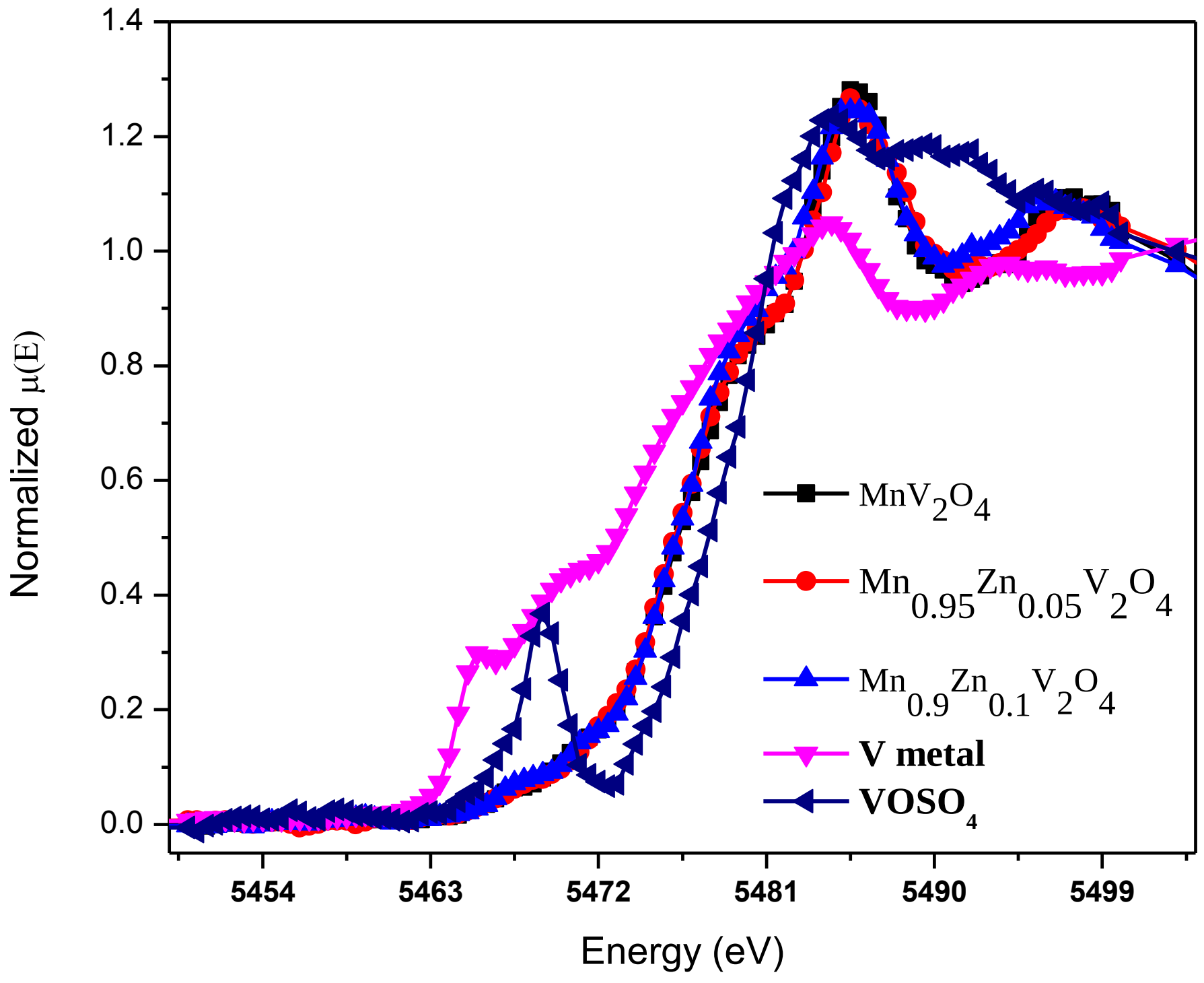

FIG.4

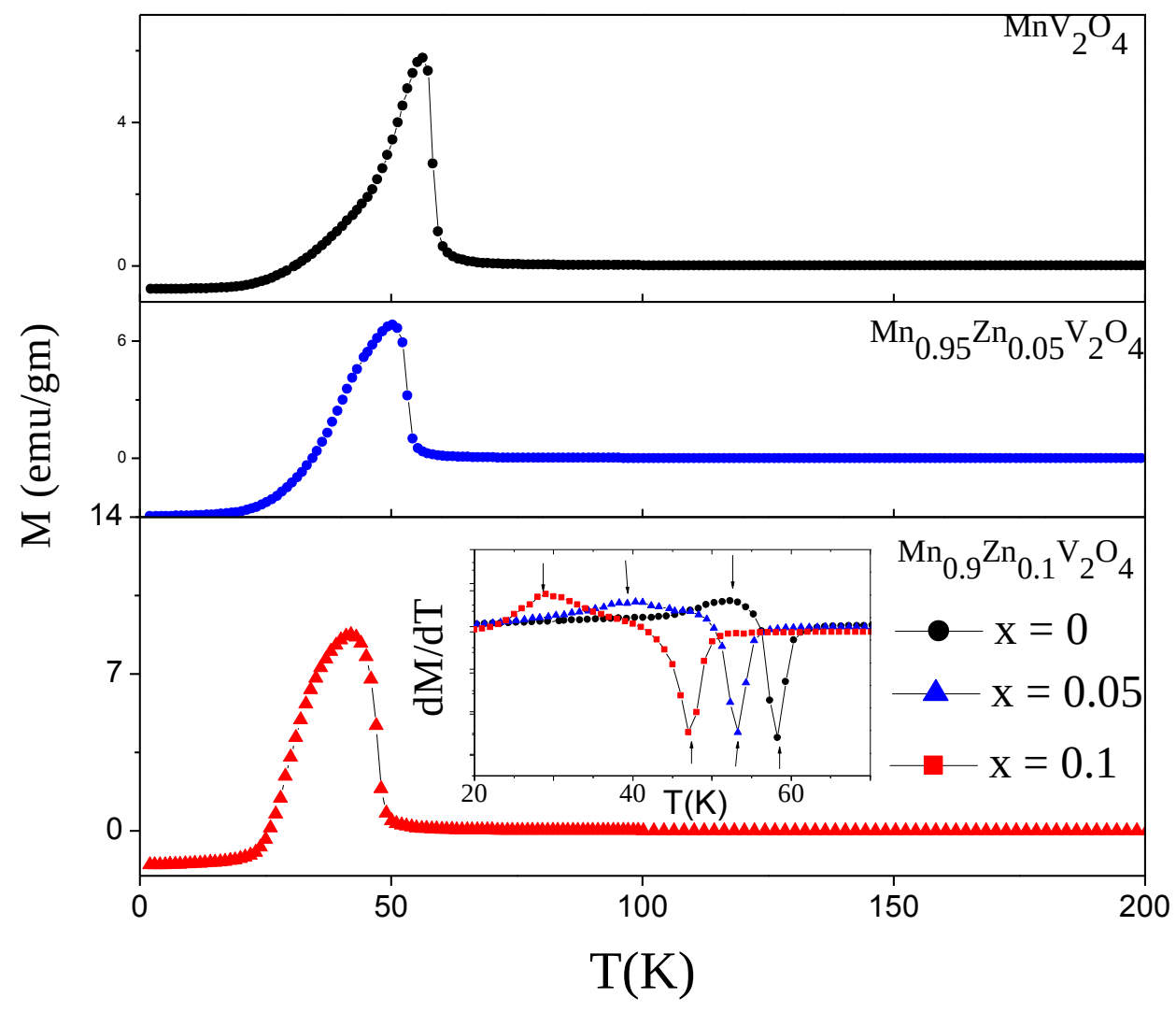

FIG. 5

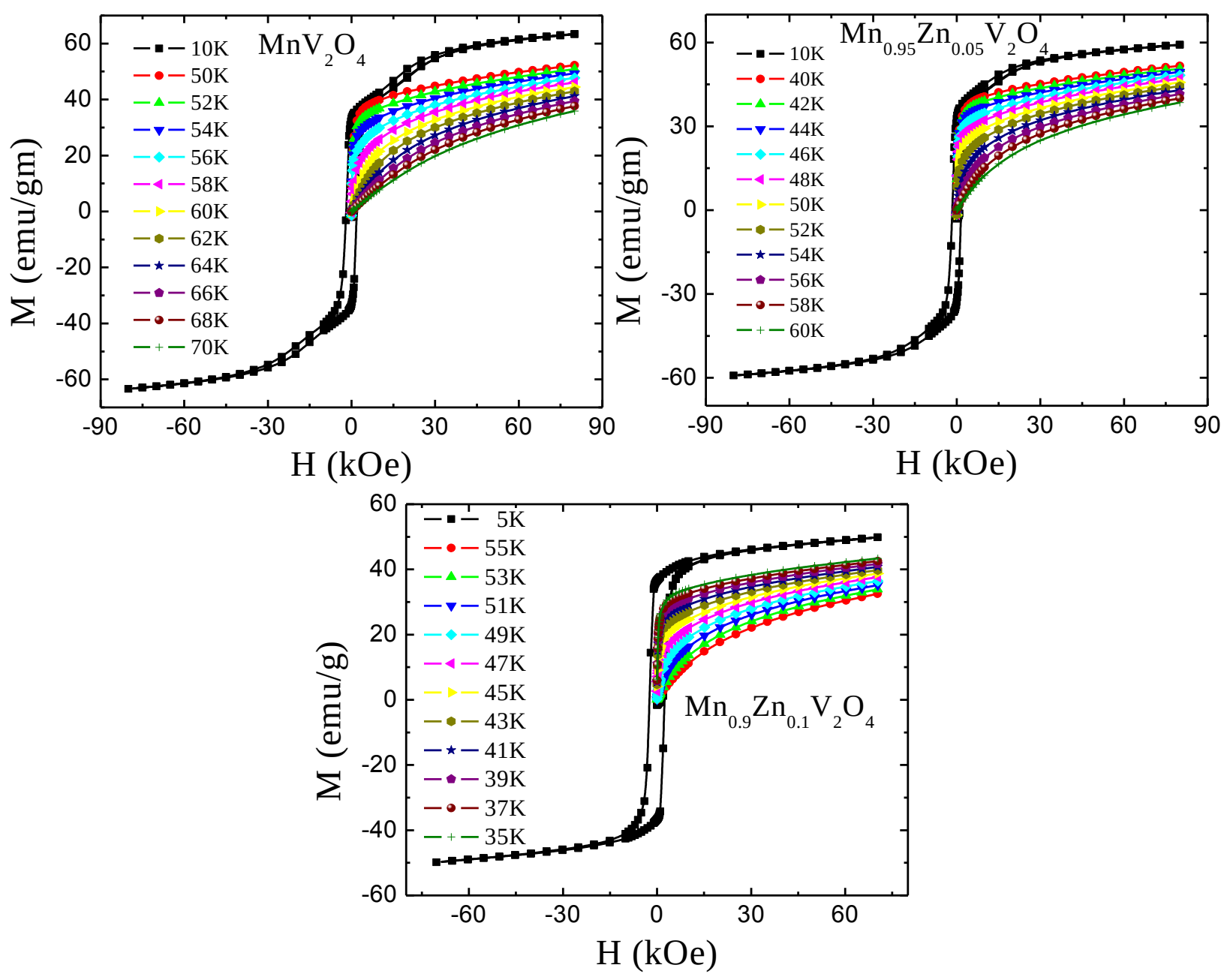

FIG.6

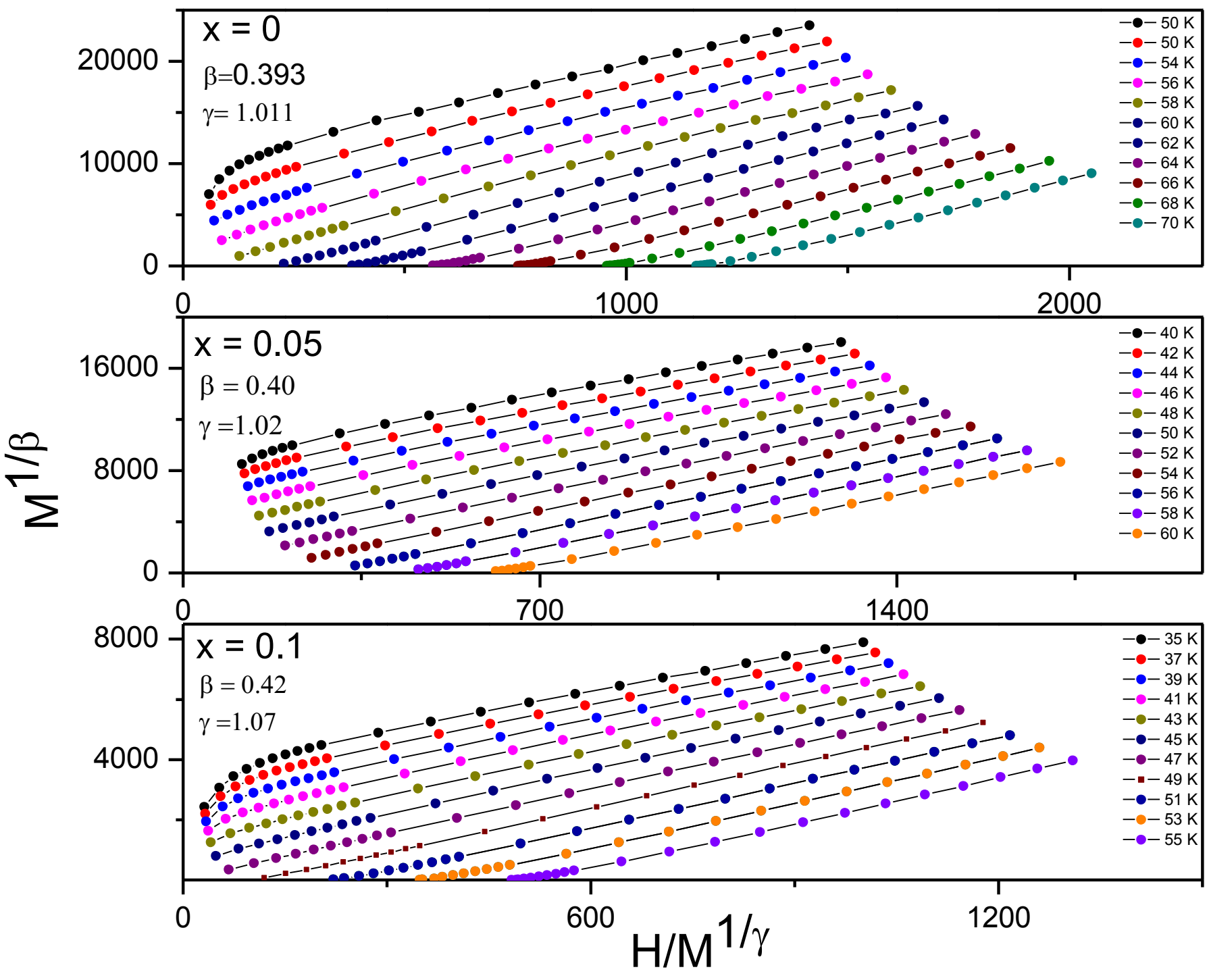

FIG.7

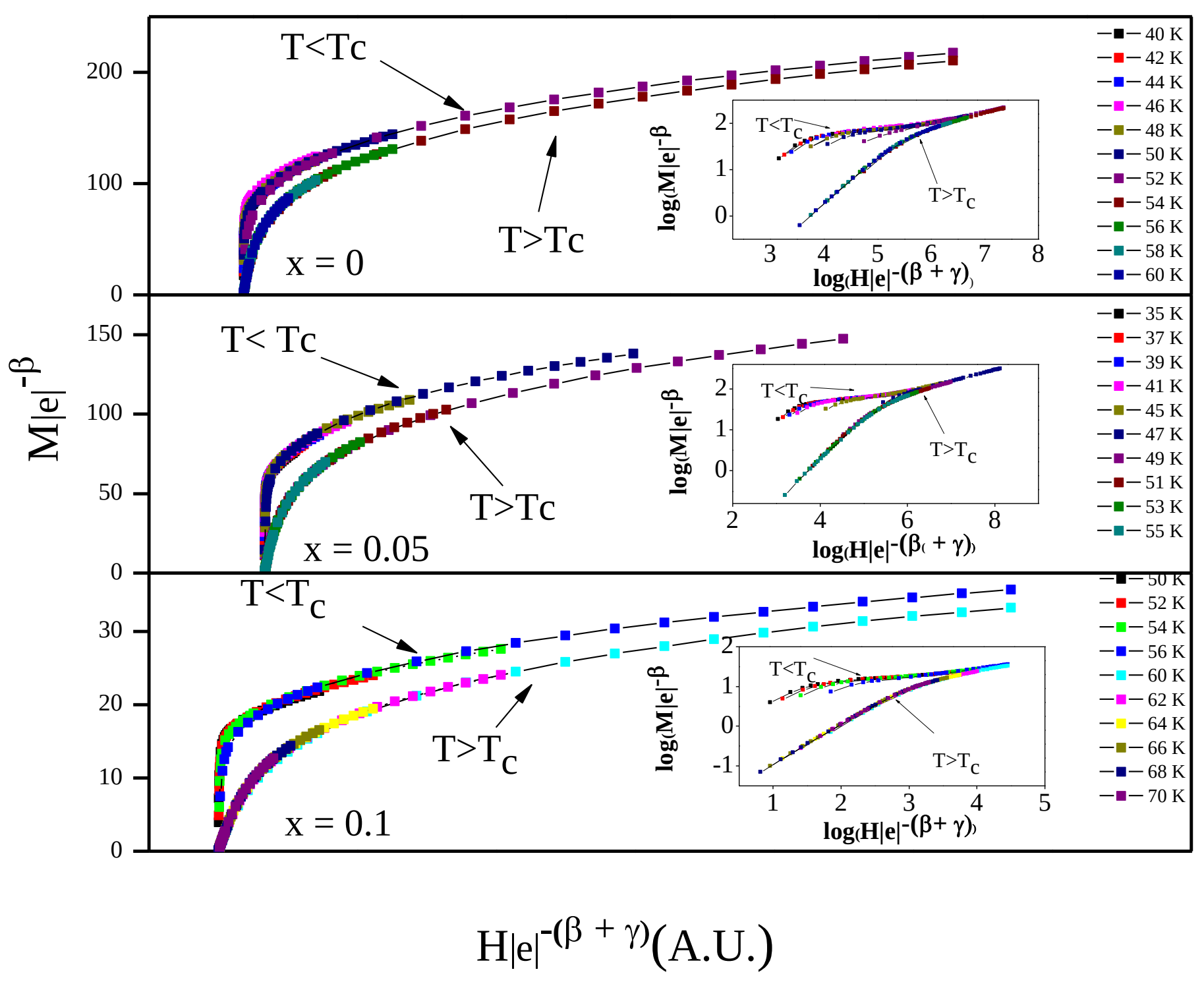

FIG.8

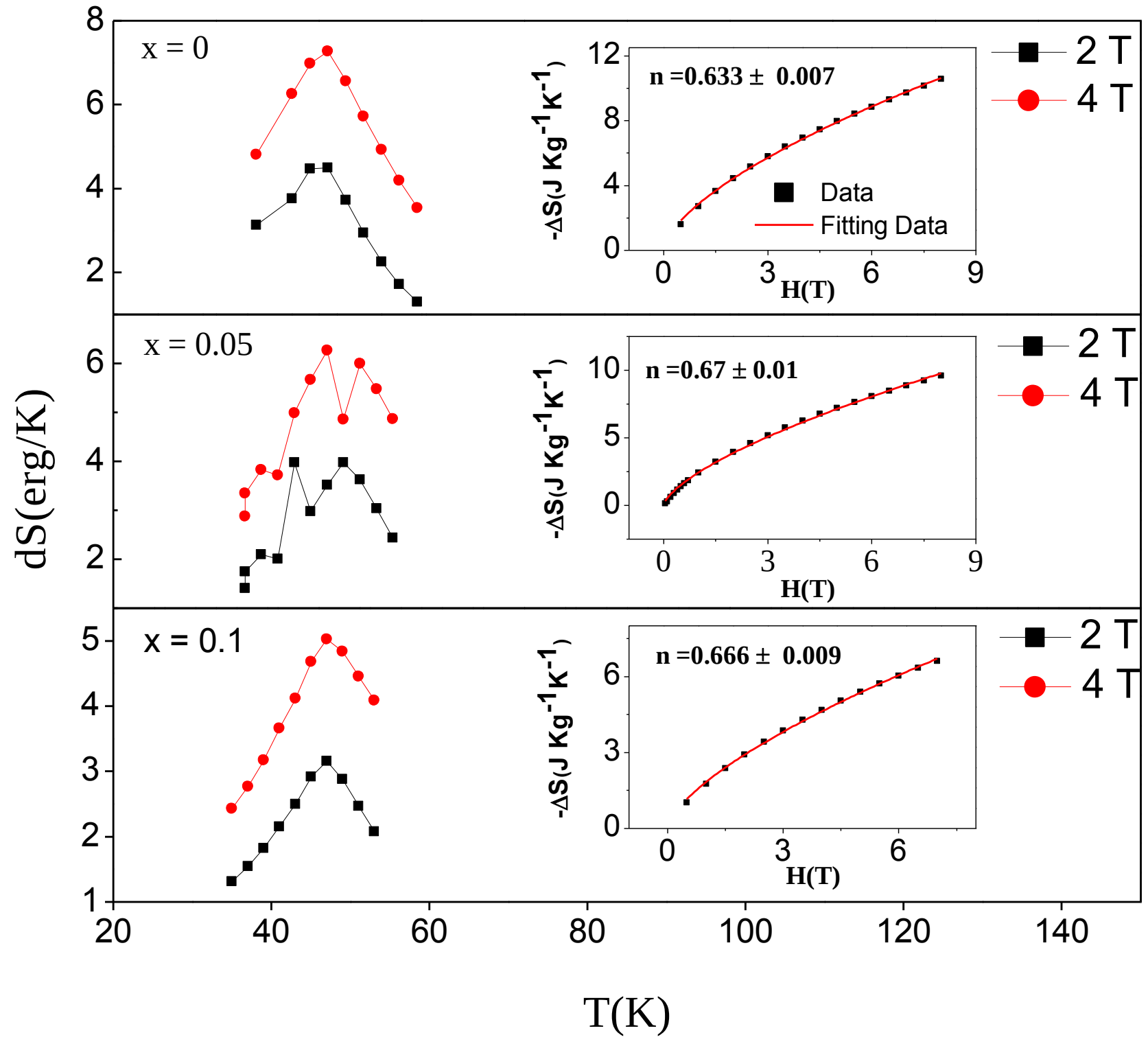

dS(erg/K)
T(K)
x = 0
x = 0.05
x = 0.1
2 T
4 T
n =0.633 ± 0.007
n =0.67 ± 0.01
n =0.666 ± 0.009
Data
Fitting Data
-ΔS(J Kg$^{-1}$K$^{-1}$)
H(T)

FIG. 9